\documentclass[12pt]{article}

\usepackage{latexsym}

\begin{document}
%%%%%%%%%%%%%%%%%%%%%%%%%%%%%%%%%%%%%%%%%%%%%%%%%%%%%%%%%%%%%%%%%%%

\title{ Cold scalar-tensor black holes coupled to \\ a massless scalar 
field}

\author{
     Stoytcho S. Yazadjiev \thanks{E-mail: yazad@phys.uni-sofia.bg}\\
{\footnotesize  Department of Theoretical Physics,
                Faculty of Physics, Sofia University,}\\
{\footnotesize  5 James Bourchier Boulevard, Sofia~1164, Bulgaria }\\
}

\date{}

\maketitle

\begin{abstract}
New four-dimensional black hole solutions of Brans-Dicke equations
with a negative constant $\omega$, coupled to a massless scalar
field, are presented. The temperature of these black holes is
zero and the horizon area is infinite. An astrophysical
application is also discussed.
\end{abstract}

%%%%%%%%%%%%%%%%%%%%%%%%%%%%%%%%%%%%%%%%%%%%%%%%%%%%%%%%%%%%%%%%%%%

%\draft
\sloppy
%\scrollmode
%%%%%%%%%%%%%%%%%%%
\renewcommand{\baselinestretch}{1.3} %
\newcommand{\sla}[1]{{\hspace{1pt}/\!\!\!\hspace{-.5pt}#1\,\,\,}\!\!}
\newcommand{\db}{\,\,{\bar {}\!\!d}\!\,\hspace{0.5pt}}
\newcommand{\partb}{\,\,{\bar {}\!\!\!\partial}\!\,\hspace{0.5pt}}
\newcommand{\dsla}{\partb}
\newcommand{\eql}{e _{q \leftarrow x}}
\newcommand{\eqr}{e _{q \rightarrow x}}
\newcommand{\ite}{\int^{t}_{t_1}}
\newcommand{\itz}{\int^{t_2}_{t_1}}
\newcommand{\itd}{\int^{t_2}_{t}}
\newcommand{\lfrac}[2]{{#1}/{#2}}
\newcommand{\dV}{d^4V\!\!ol}
\newcommand{\ben}{\begin{eqnarray}}
\newcommand{\een}{\end{eqnarray}}
\newcommand{\la}{\label}

%%%%%%%%%%%%%%%%%%%

Scalar-tensor theories of gravity are considered as the most
natural generalizations of general relativity. From theoretical
point of view one should mention that specific scalar-tensor
theories arise naturally as a low energy limit of string theory.
One of the most interesting aspects of scalar-tensor theories is
the possible existence of black holes different form those in
general relativity. Campanelli and Lousto \cite{CL} pointed out
that, in Brans-Dicke theory with a negative coupling constant
$\omega$, there exist  solutions possessing all black holes
properties but with horizons of infinite area (type B black
holes). Let us note that that Brans-Dicke theory with $\omega<0$
is also worth to study for more than its intrinsic theoretical
interest. For sufficiently negative $\omega$, the theory passes
all the experimental constrains as does for possitive $\omega$
\cite{Will}.

Later on the work of Campanelli and Lousto  was generalized in
\cite{BCEF1} where it was shown, in the framework of a general
class of scalar-tensor theories, that nontrivial black hole
solutions can exist for the coupling function $2\omega(\Phi)+
3<0$. These black holes solutions are divided into two classes:
class $B1$, where horizons are attained by infalling particles in
a finite proper time, and class $B2$, for which this proper time
is infinite. The structure and stability of the scalar-tensor
black holes were investigated in \cite{BCCF}. Charged
scalar-tensor black holes were studied in \cite{BCEF2} where it
was shown that they exist for  anomalous versions of the
scalar-tensor theories with a negative kinetic term in the
lagrangian. The temperature of these black holes is zero and the
horizon area is (in most cases) infinite.

Thermodynamics of the black holes with infinite horizon area was
considered in \cite{Z}.

Stationary, axisymmetric black holes in Bans-Dicke theory with
$-5/4 \le \omega < -3/2$ were discussed in \cite{KIM}. These
black holes are characterize with zero Hawking temperature and
finite horizon area.

The aim of the present work is to present new type scalar-tensor
black hole solutions of Brans-Dicke equations with $2\omega +3<0$,
coupled to a real massless scalar field.

The action for the scalar-tensor gravity in the presence of a
real massless scalar filed $\sigma$ is:

\begin{equation}
S = {1\over 16\pi } \int d^4x\sqrt{-g} \left(\Phi R - {\omega
\over \Phi} g^{\mu\nu}\partial_{\mu}\Phi \partial_{\nu}\Phi -
2g^{\mu\nu}\partial_{\mu}\sigma \partial_{\nu}\sigma \right)
\end{equation}

The field equations are given by:

\begin{eqnarray} \label{BDFE}
R_{\mu\nu} - {1\over 2}g_{\mu\nu}R &=& {1 \over \Phi} T_{\mu\nu} +
{\omega \over \Phi^2} \left(\partial_{\mu}\Phi
\partial_{\nu}\Phi  - {1\over 2} g_{\mu\nu}\partial_{\alpha}\Phi 
\partial^{\alpha}\Phi  \right)  \\
&+&  {1\over \Phi}\left(\nabla_{\mu}\nabla_{\nu}\Phi -
g_{\mu\nu}\nabla_{\alpha}\nabla^{\alpha}\Phi  \right), \\
\nabla_{\alpha}\nabla^{\alpha}\Phi &=& {T \over 3 + 2\omega}, \\
\nabla_{\alpha}\nabla^{\alpha}\sigma &=& 0 ,
 \end{eqnarray}

where

\begin{equation} \label{EMT}
T_{\mu\nu} = 2\partial_{\mu}\sigma \partial_{\nu}\sigma -
g_{\mu\nu}\partial_{\alpha}\sigma \partial^{\alpha}\sigma
\end{equation}

and $T= g^{\mu\nu}T_{\mu\nu}$.

We were able to find\footnote{This class of exact solutions can
be obtained by using solution generating techniques developed in
\cite{Y}.} the following static, spherically symmetric class of
exact solutions of Eq.(\ref{BDFE})-Eq.(\ref{EMT}) for $2\omega +
3 < 0$:

\begin{eqnarray}
ds^2 &=& \left({1 -\beta^2 f^b(r)\over 1-\beta^2 }\right)^2 \left(
-f^{a-b}(r)dt^2  + f^{ - a-b}(r)dr^2 +
f^{1 -a - b}(r) r^2 d\Omega^2 \right), \\
\Phi^{-1}(r) &=& \left({1 -\beta^2 f^b(r)\over 1-\beta^2 }\right)^2 
f^{-b}(r), \\
\sigma(r) &=&  \pm \beta \mid 3+ 2\omega\mid^{1/2} {1 -
f^{b}(r)\over 1 - \beta^2 f^{b}(r)}
\end{eqnarray}

where

\begin{equation}
f(r) = 1 -{2\lambda\over r}.
\end{equation}

The found class depends on three essential parameters $\lambda$,
$\beta$ ($\beta^2 <1$) and $a$. The parameters $b$ and $a$ are
related by

\begin{equation}
(2\omega + 3)b^2 = 1 - a^2
\end{equation}

and therefore $a^2 \ge 1$. Here we shall consider the solutions
with $\lambda >0$.

Let us note that the usual vacuum Brans-Dicke solution (with
$2\omega +3 <0$) is recovered for $\beta=0$.

Black hole solutions with a metric

\begin{equation}
ds^2 = g_{00}(r)dt^2 + g_{11}(r)dr^2 + g_{22}(r)d\Omega^2
\end{equation}

are selected by the following criteria \cite{BCEF1}:

\begin{enumerate}
\item There exists a Killing horizon: at some $r=r_{*}$,
$g_{00}(r_{*})=0$;

\item The integral $\int \left({g_{11}\over\mid g_{00}\mid}\right)^{1/2}
dr$ $\to$ $\infty$ as $r\to r_{*}$ (invisibility of the horizon
for an observer at rest);

\item The Hawking temperaure $T_{H}$ is finite;

\item The invariants ${\cal K}_{1}=R^2$, ${\cal 
K}_{2}=R_{\mu\nu}R^{\mu\nu}$,
${\cal K}_{3} = R_{\mu\nu\alpha\beta}R^{\mu\nu\alpha\beta}$ are
finite at $r=r_{*}$.

\end{enumerate}

It can be checked that all criteria are satisfied by the solutions
with

\begin{equation}
1 < a ,  \; \;  0 < b , \;  \; 2 - a \le b < a.
\end{equation}

Therefore, these solutions describe scalar-tensor black holes
with a regular horizon at $r=r_{*}=2\lambda$. These black holes
are cold since $T_{H}=0$ and have a horizon with an infinite area
(i.e. type B black holes). It can be shown that the infalling
particles reach the horizon for finite proper time when $0<b<1$
(type B1 black holes) and this time is infinite when $b\ge 1$
(type B2 black holes). It is worth also noting that the scalar
field $\sigma$ is regular on the horizon. Following the approach
presented in \cite{Z} it can be shown that our  black holes  are
characterized  with zero entropy (${\cal S}=0$).

In this work we presented new type black hole solutions in
Brans-Dicke theory with a negative coupling parameter $\omega$ :
black holes coupled to a massless scalar field. These black hole
solutions are interesting not only from a pure theoretical point
of view. They, as well as the vacuum scalar-tensor black holes,
could be of astrophysical relevance and some astrophysical
applications were discussed in \cite{CL}. Here we would like to
point out another astrophysical application. Gravitational lensing
of the scalar-tensor black holes in the strong field regime could
be quite different from that of the general relativistic black
holes and could exhibit some new features. In fact, the
gravitational lensing of objects (naked singularities) with a
space-time metric similar to that of the scalar-tensor black
holes was investigated by Virbhadra and Ellis in \cite{VE1}. They
found that the gravitational lensing of the strongly naked
singularity is qualitatively different from that of a
Schwarzschild black hole \cite{VE2}.  However, while the metric
studied in \cite{VE1} is characterized with a parameter $a<1$ (in
our notations, and $\nu <1$ in the notations of \cite{VE1}), the
metrics of scalar-tensor black holes have $a>1$. This could lead
to qualitatively new effects. So, the strong field gravitational
lensing may be used to distinguish between the scalar-tensor and
general relativistic black holes. Of course, this question needs
more careful investigation .

\bigskip

\bigskip

\noindent{\Large\bf Acknowledgments}

\vskip 0.3cm

I would like to thank V. Rizov for reading the manuscript. This
work was partially supported by Sofia University Grant No3429.

\end{document}